\documentclass[11pt,a4paper]{article}

\usepackage{authblk}
\usepackage{amssymb}
\usepackage{amsthm}

\usepackage{graphicx}        

\begin{document}

\title{Experimental Analysis of Two-Dimensional Pedestrian Flow in front of the Bottleneck}
\author{Marek Buk\'a\v cek, Pavel Hrab\'ak, Milan Krb\'alek}
\affil{Faculty of Nuclear Sciences and Physical Engineering, Czech Technical University in Prague, Czech Republic}

\maketitle

\begin{abstract}
This contribution presents experimental study of two-dimensional pedestrian flow with the aim to capture the pedestrian behaviour within the cluster formed in front of the bottleneck. Two experiments of passing through a room with one entrance and one exit were arranged according to phase transition study in \cite{Ezaki2012}, the inflow rate was regulated to obtain different walking modes. By means of automatic image processing, pedestrians' paths are extracted from camera records to get actual velocity and local density. Macroscopic information is extracted by means of virtual detector and leaving times of pedestrians. The pedestrian's behaviour is evaluated by means of density and velocity. Different approaches of measurement are compared using several fundamental diagrams. Two phases of crowd behaviour have been recognized and the phase transition was described. 
\end{abstract}

\section{Introduction}

One of the main impacts of pedestrian behavior's study is the ability to optimize the infrastructure. Using some intervention, the capacity of given zone (building, public area, transportation hub, etc.) can be increased, i.e. more people can pass this zone with higher velocity and lower number of conflicts \cite{Helbing2000}, \cite{Schadschneider2009}. 

Many studies deal with certain aspects of pedestrian motion \cite{Boltes2011}, \cite{Jelic2012}, \cite{Seyfried2010}, \cite{Seyfried2010a}, \cite{Was2010}, \cite{Zhang2010}. Specific simulation tools are often supported by experimental data analyses \cite{Federici2012}, \cite{Georgoudas2011}, \cite{Hrabak2012}, \cite{Was2011}.

This article focuses on the description of the system as a whole. In particular we focus on the transition from the free to the congestion phase. This study is compared with the actual states of individual pedestrians in the system as explained below.

The density in the simplest form represents the number of pedestrians in fixed area (referred to as $\rho_A$) \cite{Schadschneider2010}. As mentioned in \cite{Steffen2010}, this quantity can be understood locally as well. More precisely, the density $\rho_{\alpha}$ in the neighborhood of pedestrian $\alpha$ corresponds to inverse value of his space consumption, i.e., the area of his Voronoi cell. 

The flow is defined as the number of persons, who crossed given intersection during one time unit. The flow through given area (e.g. detector area) can be evaluated from the number of pedestrians $N^{+}_t$, who entered into the monitored area A during $\langle t, t+\Delta t \rangle$, i.e., $J_{A}(t) = \frac{N^{+}_t}{\Delta t}$. The specific flow is related to uniform corridor width, therefore $J_{A}^{s}(t) = \frac{J_{A}(t)}{d}$, where $d$ represents a width of given corridor.

Using hydrodynamic approach, the density, velocity, and flow can be related by formula
\begin{equation}
J(\rho) = \rho v(\rho).
\end{equation}

Both, the relations $v = v(\rho)$ or $J = J(\rho)$, are referred to as fundamental diagrams (FD) \cite{Schadschneider2009} and are used to illustrate the essential behaviour of the system.

In this article, we distinguish two variants of FD:
\begin{itemize}
\item Area FD -- The system is observed through defined area to extract the dependence $J_A(t) = J_A(t,\rho_A)$. Data cumulation over long interval $T$ allows to observe pedestrian behavior under different conditions, e.q., the density limits, which characterize the phase transition. 
\item Individual FD -- For each pedestrian $\alpha$ the dependence $v_{\alpha}(t) = v_{\alpha}(t,\rho_{\alpha})$ is evaluated. By observing pedestrians under different conditions, one can identify the conditions preceding the decrease of velocity. 
\end{itemize}

Considering a room with one entrance and one exit, the following observations can be made. With low inflow, pedestrians can exit freely and move with maximal velocity. By increasing the inflow rate, the number of pedestrians in the room increases. Therefore, a cluster is created in front of the exit, which means that a pedestrian is forced to slow down and integrate into the cluster before leaving the room. 

The motion of chosen pedestrian $\alpha$ can be classified according to his/her velocity and local density into three states:
\begin{itemize}
\item Free state -- the pedestrian does not react to other pedestrians and moves with his desired velocity $v_{\alpha}^{0}$. This state is characterize by low density $\rho_{\alpha}$ and high velocity $v_{\alpha}$.
\item Synchronized state -- the pedestrian motion is highly synchronized with pedestrians in his surrounding due to high density. This state is characterize by high density $\rho_{\alpha}$ and low velocity $v_{\alpha}$.
\item Transition state -- The transition between free and synchronized state is characterized by low density $\rho_{\alpha}$ and low velocity $v_{\alpha}$ due to the anticipation or long reaction time, respectively.
\end{itemize}

Analogously, the phase of the entire system can be classified:
\begin{itemize}
\item Free phase -- no cluster is formed in front of the exit, therefore the majority of pedestrians in the system is in free state
\item Congested phase -- stable cluster in front of the exit is formed, which leads to permenent significant ratio of pedestrian in the synchronized state.
\item Metastable phase -- unstable cluster arises and disappears, the ratio of pedestrian in the synchronized state changes in time from low to significant. 
\end{itemize}

\section{Experiment}

\begin{figure}[t]
\begin{center}
\includegraphics[width=0.47\textwidth]{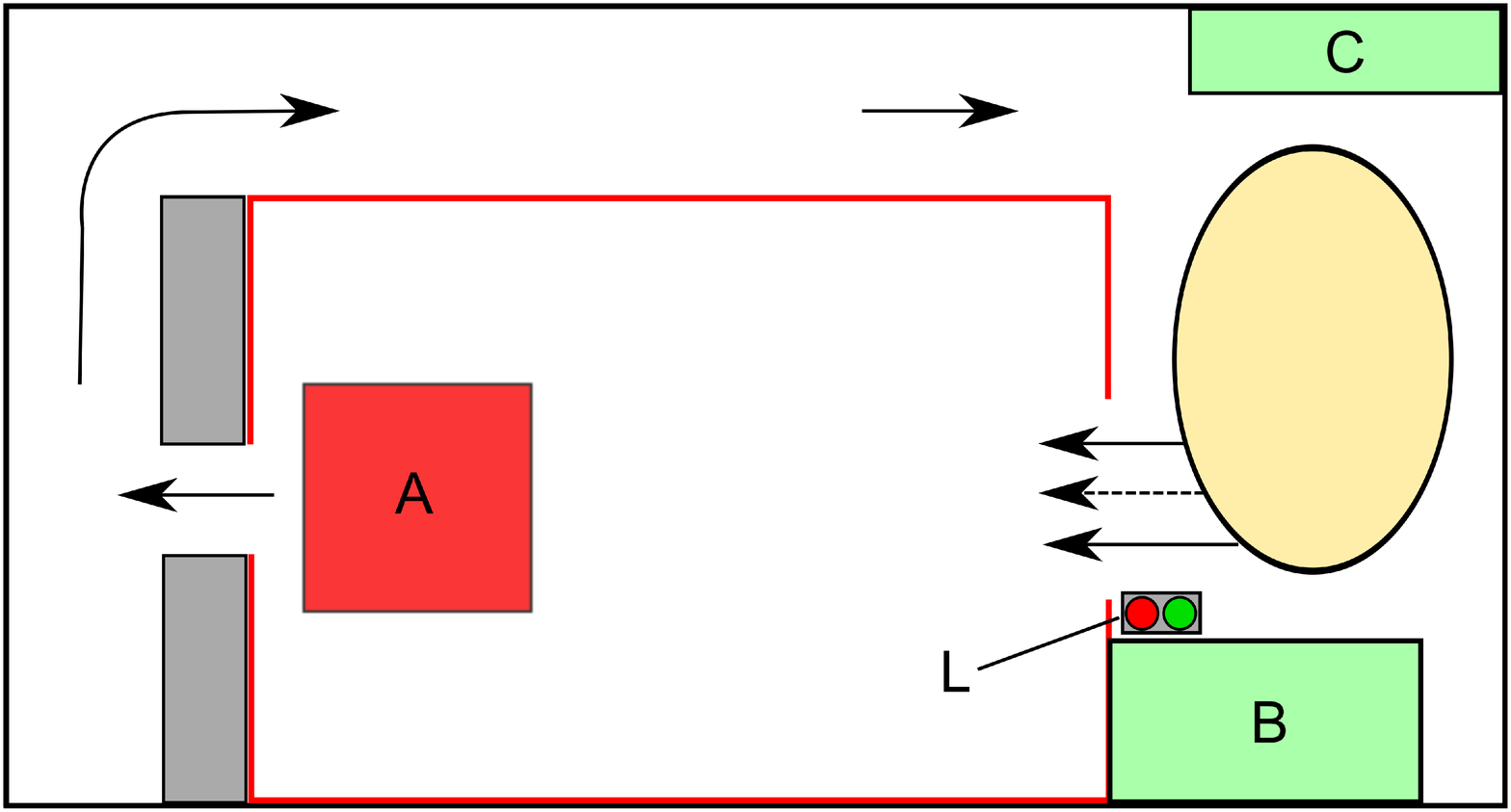}
\includegraphics[width=0.47\textwidth]{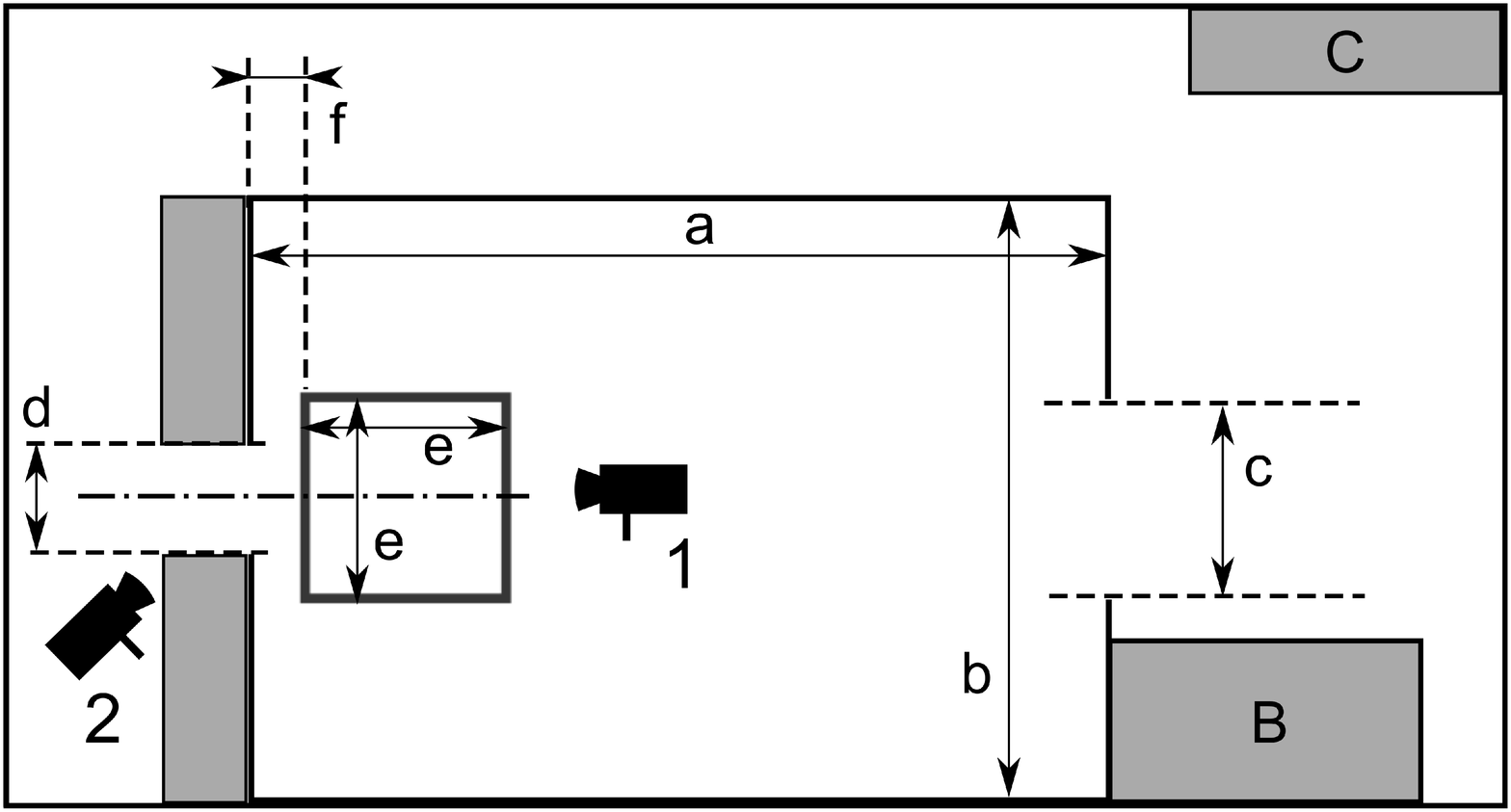}
\end{center}
\caption{Left: the schematic view of the experiments (area A represents virtual detector, B technical background, C coffee corner, L refers to traffic lights). Right: Room geometry and position of cameras. Exp.1: a~=~10 m, b = 6 m, Exp.2: a = 6 m, b = 3.5 m. Both: c = 2 m, d = 0.6 m, e = 2 m, f~=~0.5~m.}
\label{fig:plan} 
\end{figure}

To detect above mentioned phases and to analyze transition among them, a simple experiment has been designed -- a group of pedestrians passed through the room with one entrance and one exit, see Figure \ref{fig:plan}. The data samples were obtained by automatic processing of video records from two experiments, which differed by the size of the room.

\begin{figure}[t]
\begin{center}
\includegraphics[height=0.35\textwidth]{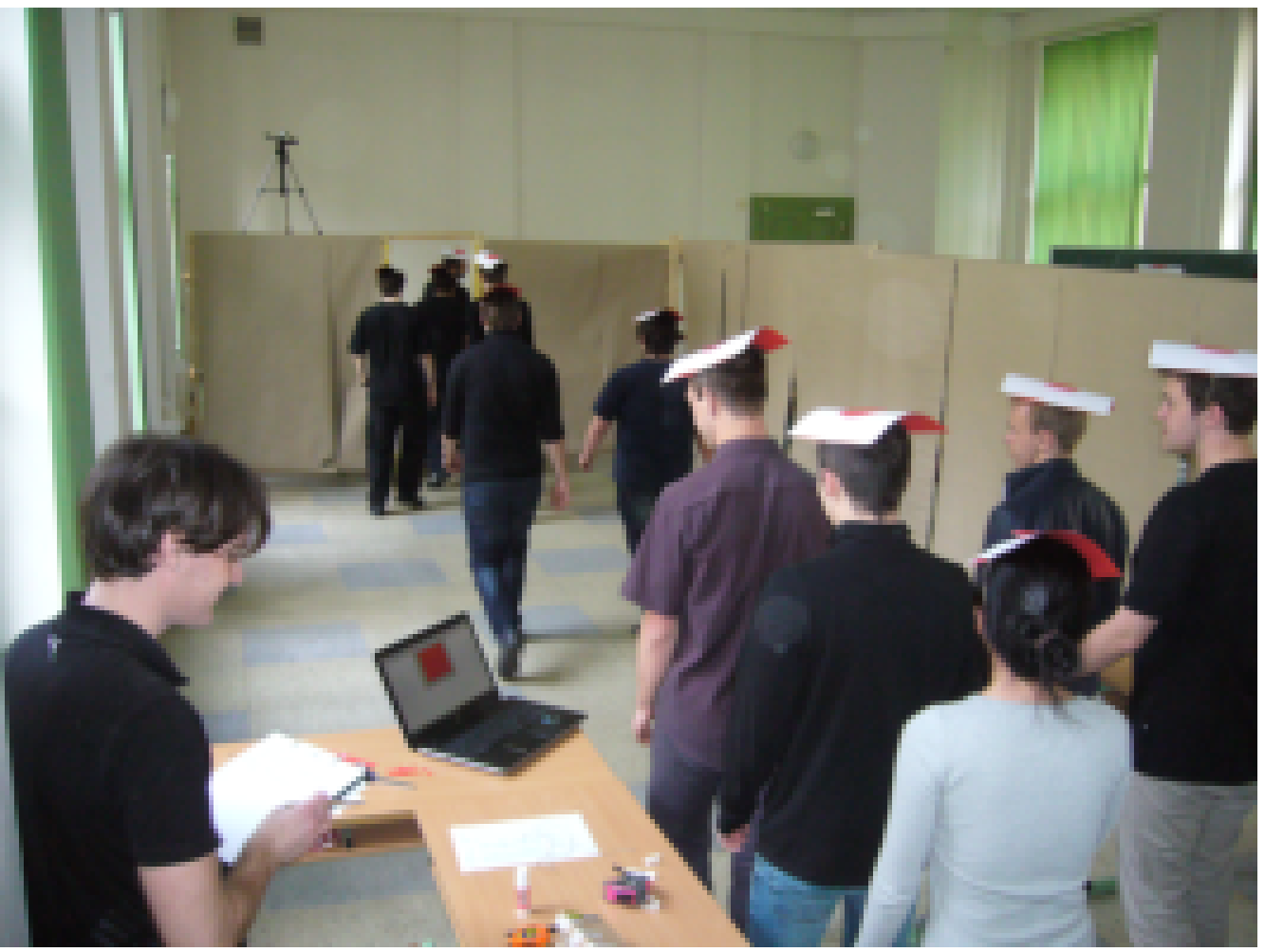}
\includegraphics[height=0.35\textwidth]{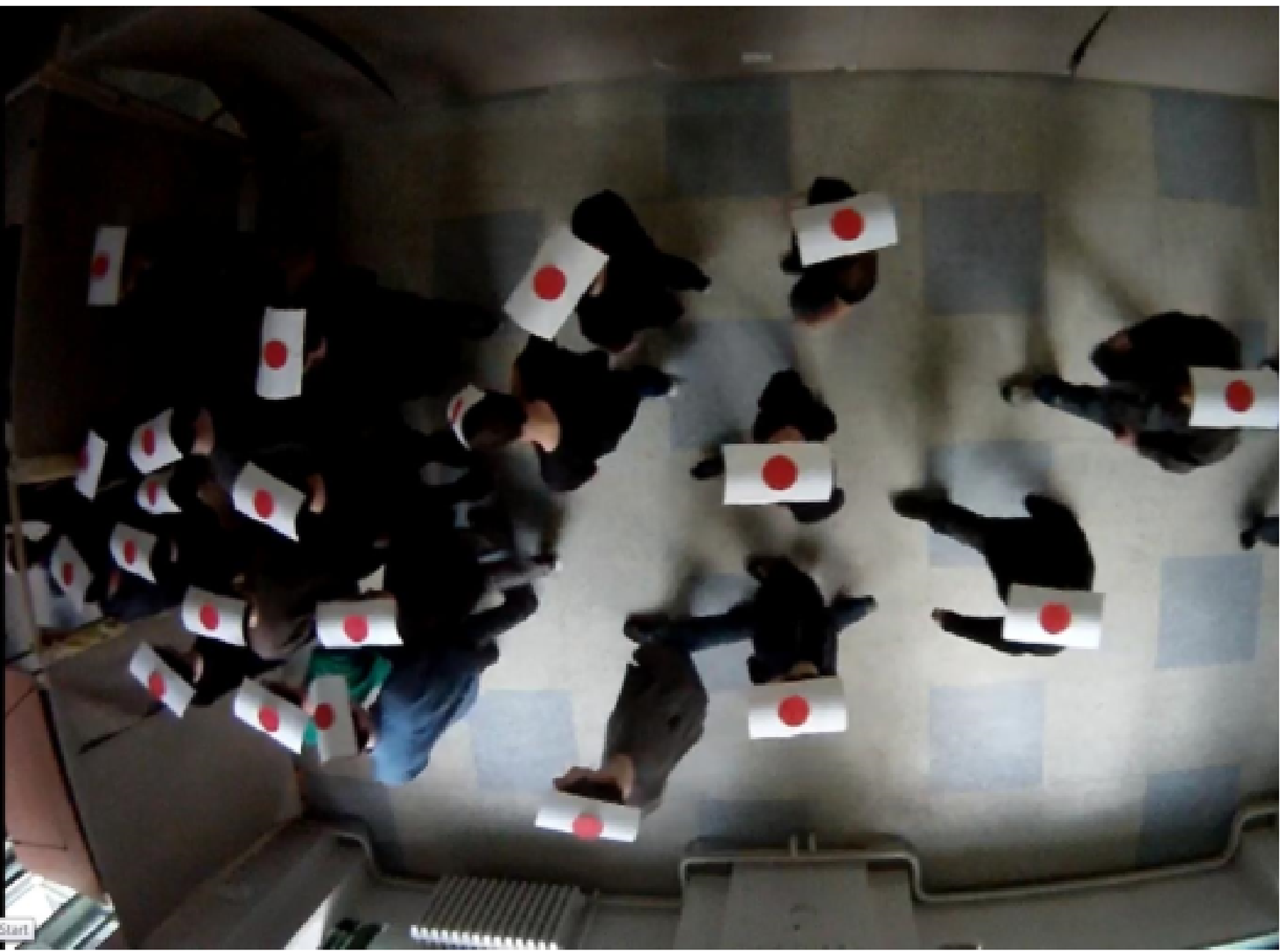}
\end{center}
\caption{The illustrating shots from the experiments.}
\label{fig:ilustrace} 
\end{figure}

To realize passing-through arrangement, an experimental room was built inside the study hall of FNSPE, see the plan in Figure \ref{fig:plan}. The walls 2 m high were made of wooden construction covered by paper. Two snapshots in Figure \ref{fig:ilustrace} visualize the design and progress of the experiment.

Two cameras were used to monitor the experiment. The main camera, which covered whole room, was fixed on the ceiling 4.5 m above the floor. The rear camera monitoring the egress of pedestrians was placed next to the exit, 2.5 m above the floor.

To control the inflow into the room, which is the crucial parameter determining the phase of the system, simple signaling device has been used. On green signal, a group of pedestrian were forced to enter the room. This green signal was altered by randomly long interval of red light, the intervals $\tau$ were generated from trimmed normal distribution: VAR$(\tau) = 1s^{2}$, E$(\tau) \in \left[ 1.2,\, 1.8\right]s$.  

Since the pedestrians were not able enter the room in headway shorter than 0.5 s, the inflow was controlled by number of entering pedestrians as well. As can be seen from Table \ref{tab:prub_exp}, this method enables the control of inflow parameter, which appears to be crucial parameter determining the free of congestion phase, as observed by means of model simulations in the \cite{Ezaki2012}, \cite{Hrabak2012}.

Nine rounds with different settings were performed in first experiment, eleven in the second one. Each of them lasted from two to ten minutes. The conditions on the input has been changed after each round, the inflow slowly raised from low values (free flow) until the state with strong congestion (for details, see Table \ref{tab:prub_exp}).
 
\begin{table}
\caption{The table describes each rounds of the experiment, following qualities are evaluated: duration of a round, number of pedestrians at input and mean input period evaluated from time series generated by traffic lights. The value "MIN" referred to the round without lights, participants were asked to enter as fast as possible. The inflow is evaluated from the period and number at input.}
\label{tab:prub_exp}
\begin{center}
\begin{tabular}{c|r|r|r|r|cc}
\multicolumn{1}{c|}{round\hspace{0.1cm}} & \multicolumn{1}{c|}{\hspace{0.1cm}duration [min:s]\hspace{0.1cm}} & \multicolumn{1}{c|}{\hspace{0.1cm}input\hspace{0.1cm}} & \multicolumn{1}{c|}{\hspace{0.1cm}period [s]\hspace{0.1cm}} & \multicolumn{1}{c|}{\hspace{0.1cm}inflow [ped/s] [s]\hspace{0.1cm}} & \multicolumn{2}{c}{\hspace{1cm} observation \hspace{1cm}} \\ \hline
1 & 10:50 & 2 & 1.78 & 1.12 & \hspace{0.1cm} & free flow, occasional delay at the exit \\
2 & 10:04 & 2 & 1.68 & 1.19 & \hspace{0.1cm} & free flow, occasional delay at the exit \\
3 & 8:29 & 2 & 1.59 & 1.26 & \hspace{0.1cm} & free flow, occasional delay at the exit \\
4 & 6:36 & 2 & 1.43 & 1.40 & \hspace{0.1cm} & metastable state \\
5 & 7:16 & 3 & 1.85 & 1.62 & \hspace{0.1cm} & cluster formation with constant size \\
6 & 6:05 & 3 & 1.69 & 1.78 & \hspace{0.1cm} & congestion \\
7 & 5:24 & 3 & 1.72 & 1.74 & \hspace{0.1cm} & congestion \\
8 & 2:55 & 3 & 1.66 & 1.81 & \hspace{0.1cm} & congestion \\
9 & 2:05 & 3 & 1.57 & 1.91 & \hspace{0.1cm} & congestion \\
\hline
1 & 6:40 & 2 & 1.60 & 1.25 & \hspace{0.1cm} & flow, occasional delay at the exit \\
2 & 7:47 & 2 & 1.61 & 1.24 & \hspace{0.1cm} & free flow, occasional delay at the exit \\
3 & 5:06 & 2 & 1.50 & 1.33 & \hspace{0.1cm} & free flow, occasional delay at the exit \\
4 & 4:15 & 2 & 1.37 & 1.46 & \hspace{0.1cm} & metastable state \\
5 & 1:52 & 2 & MIN & $\pm$ 2 & \hspace{0.1cm} & congestion \\
6 & 1:31 & 2 & MIN & $\pm$ 2 & \hspace{0.1cm} & congestion \\
7 & 8:07 & 3 & 1.76 & 1.70 & \hspace{0.1cm} & cluster formation with constant size \\
8 & 4:31 & 3 & 1.70 & 1.76 & \hspace{0.1cm} & metastable state \\
9 & 2:34 & 3 & 1.56 & 1.92 & \hspace{0.1cm} & cluster formation with constant size \\
10 & 3:23 & 3 & 1.55 & 1.94 & \hspace{0.1cm} & cluster formation with constant size \\
11 & 3:13 & 3 & MIN & $\pm$ 3 & \hspace{0.1cm} & congestion \\
\end{tabular}
\end{center}
\end{table}

\section{Data Processing}

Pedestrians were marked by red paper hats with white rim, unlike \cite{Plaue2011}. The principle of contrast color \cite{Gonzales2001} was used to detect pedestrians on each frame and their positions were recorded. Due to the width of recorded area, the fish eye deformation influenced the data. This deformation was partially suppressed by sinus transformation (see \cite{Bakstein2006} for more details).
 
Coordinates were assigned to one path of given pedestrian (ref. $\alpha$) with respect to the distance to coordinates on previous frame. Therefore the trajectory of pedestrian was reconstructed: $x_{\alpha}(t) = [x_{\alpha}^{(1)}, x_{\alpha}^{(2)}](t)$, where time is understand to be discrete: $t \in \{t_0 + n \Delta t \}$, $\Delta t$ = 1/59 s. 

As mentioned in the introduction, the local density was derived from each frame and the local velocity was be extracted from paths using central differences
\begin{equation}
v_{\alpha}^x(t) = \frac{x_{\alpha} (t + k\Delta t) - x_{\alpha} (t - k\Delta t)}{2k\Delta t}, 
\end{equation}
where $k = 5$ was used to reach sufficient smooth trajectories. Thus, the trajectory data were obtained in the form $(x_{\alpha}, \rho_{\alpha}, v_{\alpha})(t)$.

The macroscopic approach was implemented by monitoring "detector" area $A$, a virtual square 2 m $\times$ 2 m placed in front of the exit, see Figure \ref{fig:plan}. In this area, the density $\rho_A$ and flow $J_A$ was evaluated. Mean value of velocity in this area (referred to as $v_A$) was calculated by means of weighted average of pedestrians' velocity inside the area. The weight of pedestrian depends on the overlapping area of detector and his voronoi cell. These detector-area-data are of the form $(\rho_A, J_A, v_A)(t)$.

The crossectional data $J_{out}(t)$ were processed semi-automatically from the rear camera. The leaving times were determined, the outflow can be calculated using leaving times $t_{\alpha}$ headways
\begin{equation}
J_{out}(t) = \frac{n_T(t)}{T}, \qquad n_T(t) = \# \left\lbrace t_{\alpha} \in \left\langle t - \frac{T}{2},\, t + \frac{T}{2} \right) \right\rbrace,
\end{equation}
where \# denotes number of set elements.

\section{Results}

\begin{figure}[t]
\begin{center}
\includegraphics[width=0.4\textwidth]{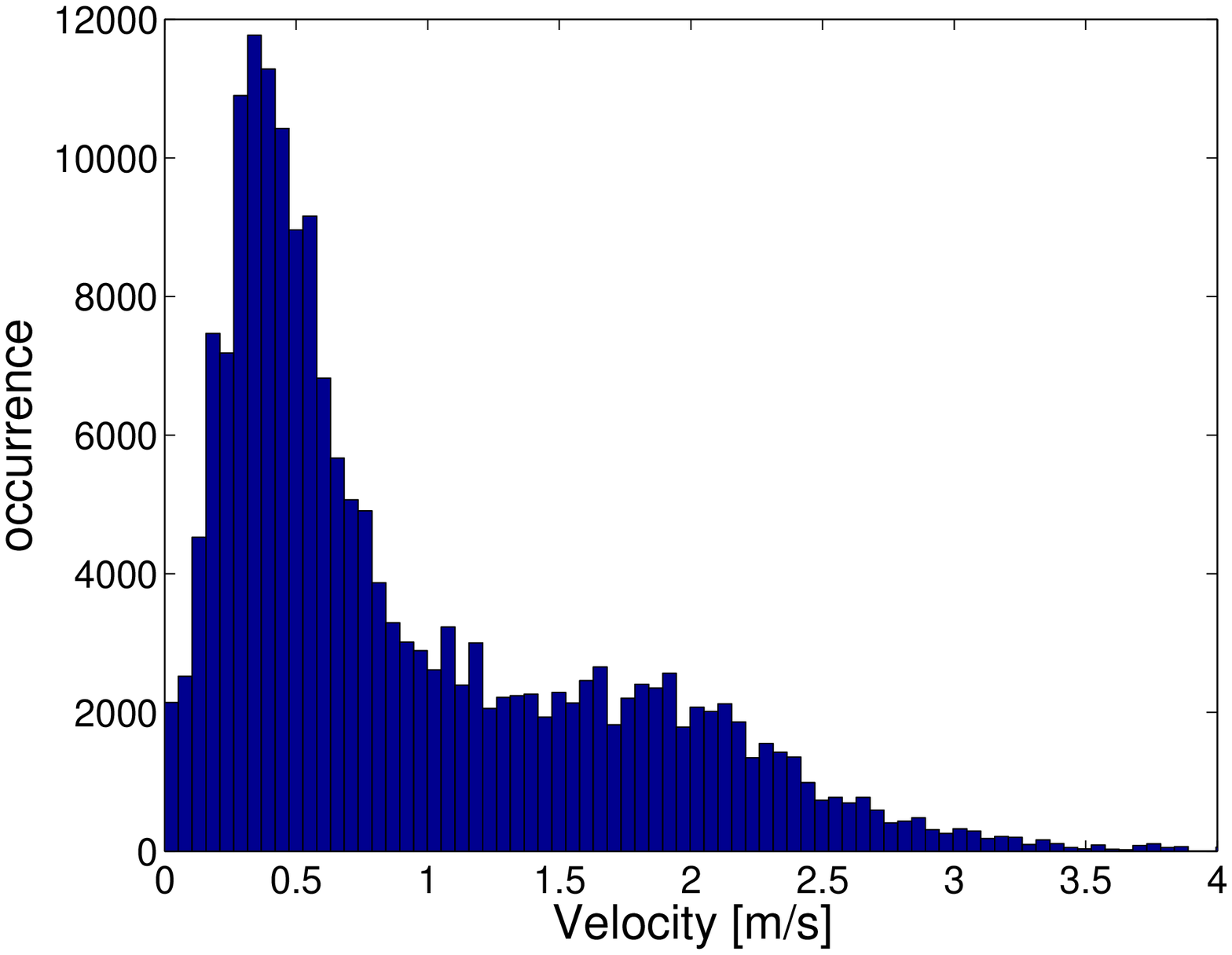}
\includegraphics[width=0.4\textwidth]{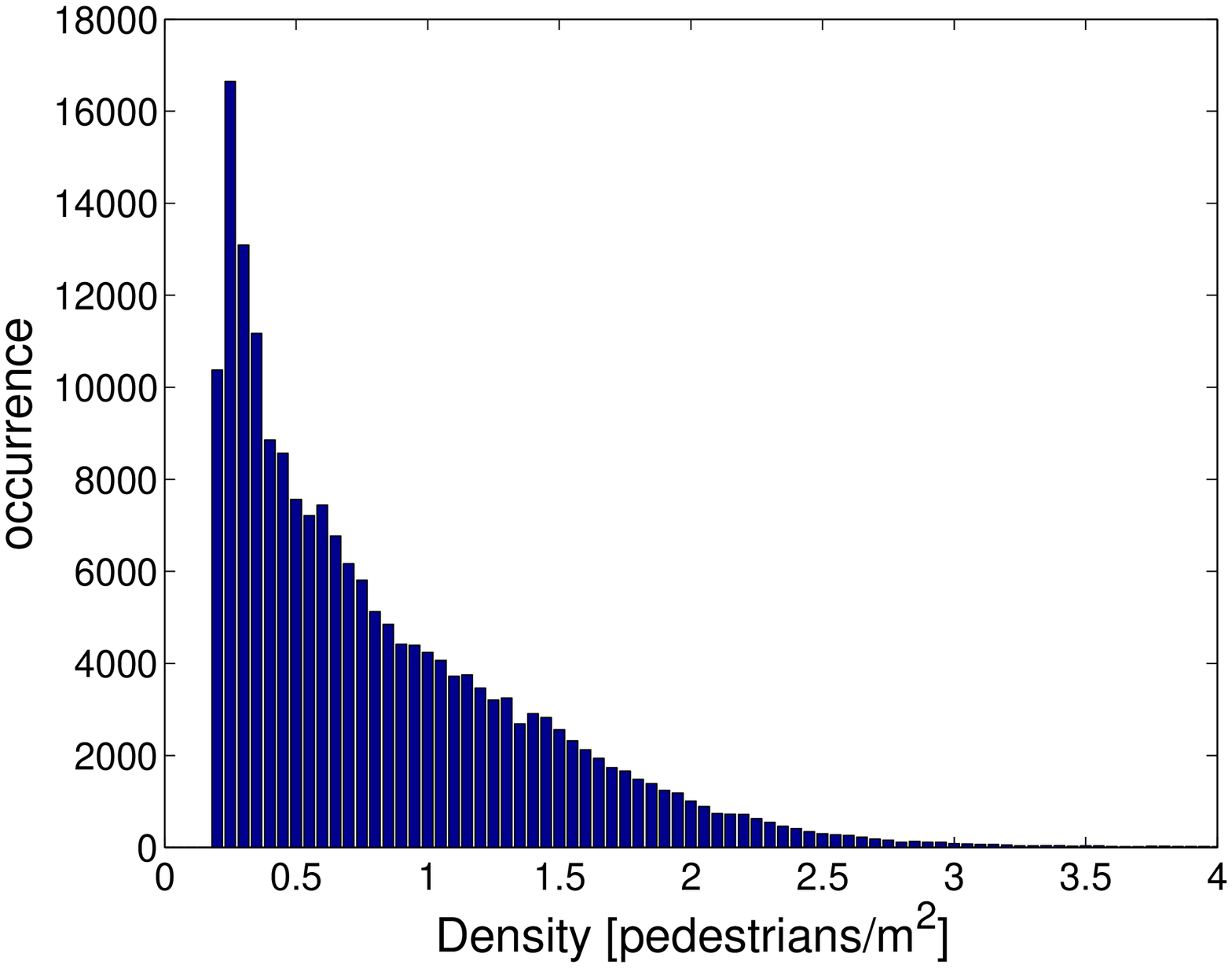}
\end{center}
\caption{Histograms of velocity and density generated from pedestrians' path data.}
\label{fig:hist} 
\end{figure}

Measured quantities can be visualized by many ways. As the first report, we show the microscopic approach. Basic review of the velocity and density is provided by their histograms -- Fig. \ref{fig:hist}. As visible, two local maxima of velovity appear: the high peak on value 0.5 m/s corresponds to synchronized state, while the wide peak on value 1.8 m/s corresponds to free motion.

\begin{figure}
\includegraphics[width=0.6\textwidth]{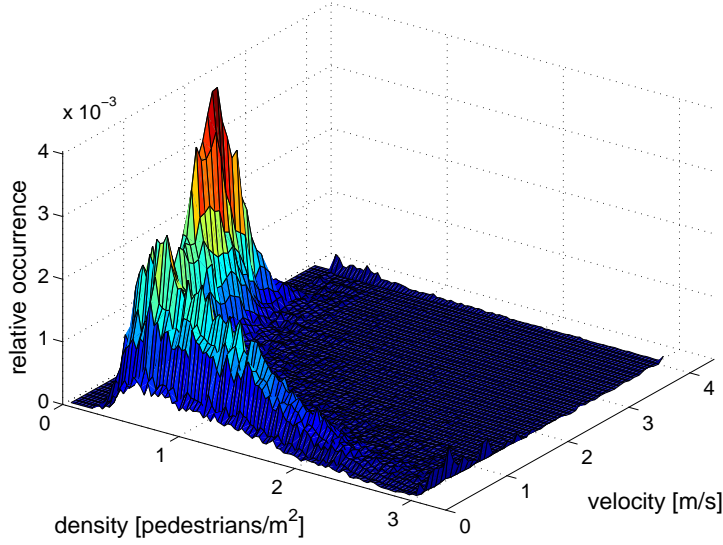}
\caption{The three dimensional fundamental diagram $v(\rho)$ generated from pedestrians' path data.}
\label{fig:3DFD} 
\end{figure}

Individual modes of motion are clearly described by the three-dimensional fundamental diagram (Fig. \ref{fig:3DFD}), regardless to their frequency of occurrence. This frequency is displayed on the $z$ coordinate.

As mentioned above, two main states were observed. Free flow occurs until the density reaches 0.3 ped/m$^{2}$. In this mode, the participants walked with velocity range 1.5 - 2.5 m/s. Conversely, when density exceeded 0.5 ped/m$^2$, congested state appeared. In this state, the velocity fluctuates between 0 and 0.7 m/s. The highest observed density was 3 ped/m$^2$. Metastable state appeared when the density occurred in the interval 0.3 -- 0.5 ped/m$^2$.

These critical values of density are significantly lower than in \cite{Zhang2010}. Such a different behavior is probably caused by two dimensional nature of investigated movement. Pedestrians slow down due to the anticipation of side conflicts.

\begin{figure}[t]
\begin{center}
\includegraphics[width=0.45\textwidth]{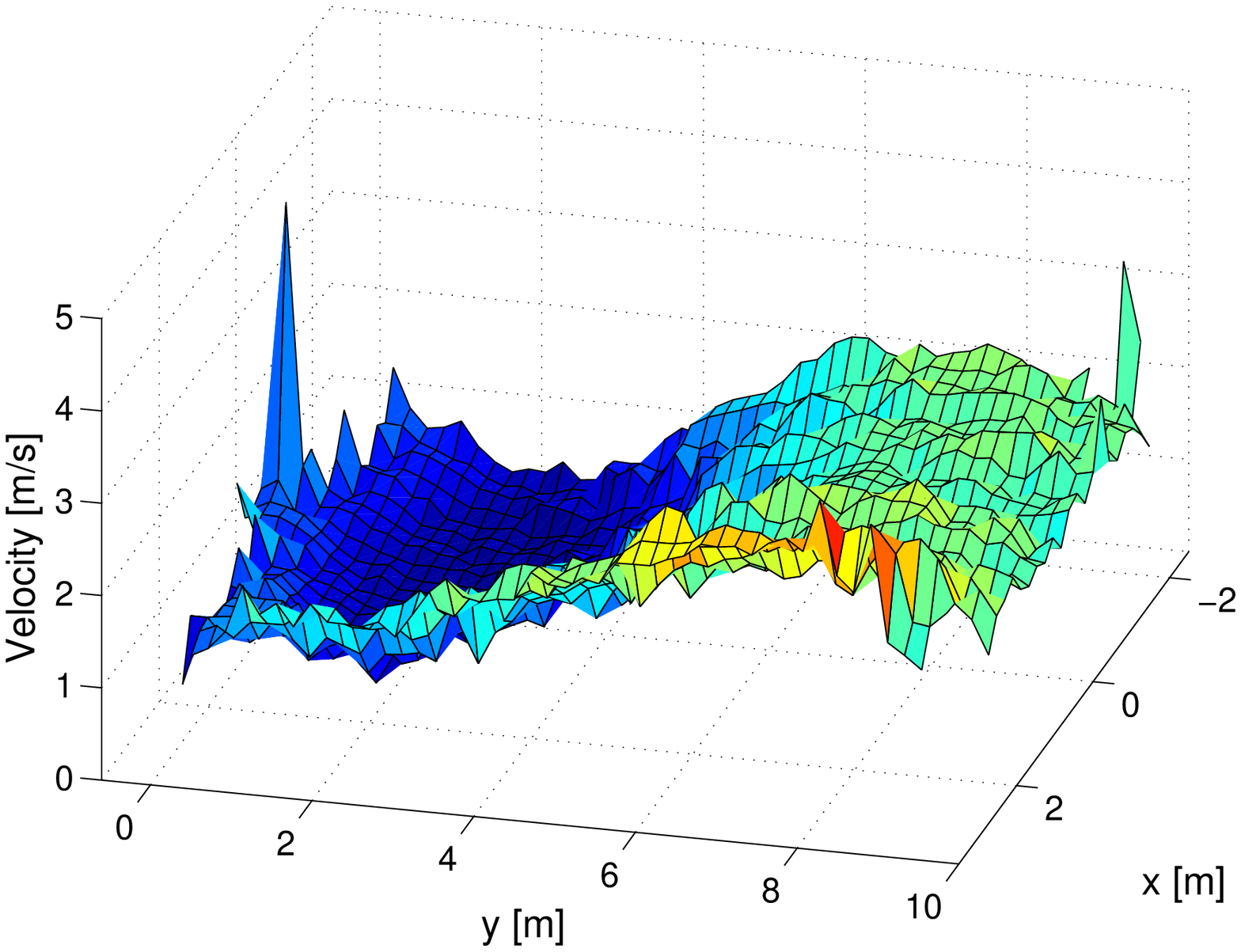}
\includegraphics[width=0.45\textwidth]{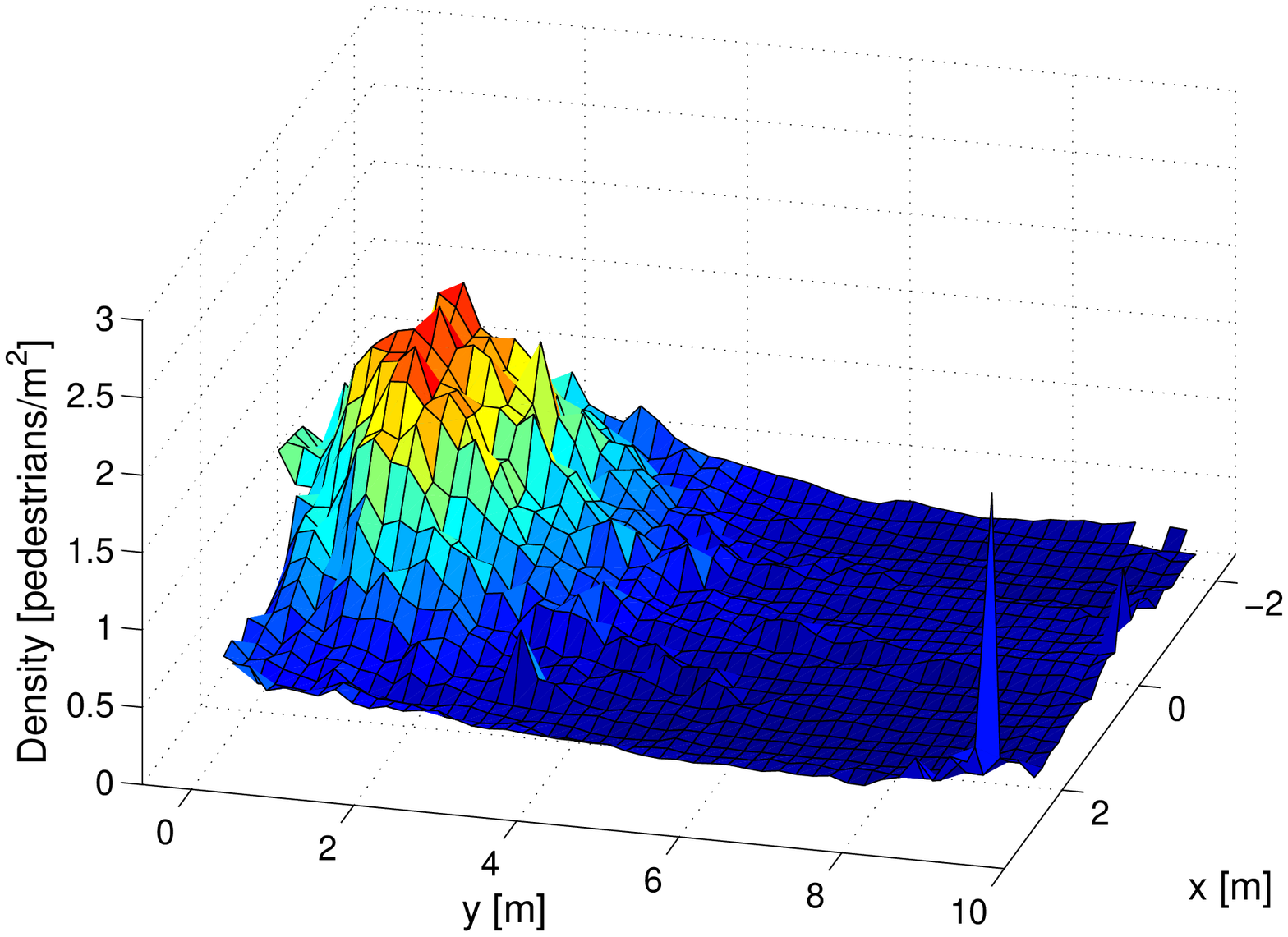}
\end{center}
\caption{The distribution of velocity and density evaluated from pedestrians' path data.}
\label{fig:E2_vel_hust} 
\end{figure}

The phase transition can be monitored by observing the distribution of velocity and density in the room (Figure \ref{fig:E2_vel_hust}). One can see that the distance between maximal velocity and high density, which represents the process of transition from free motion to the congested state, reaches 3 m.

\begin{figure}
\includegraphics[width=0.5\textwidth]{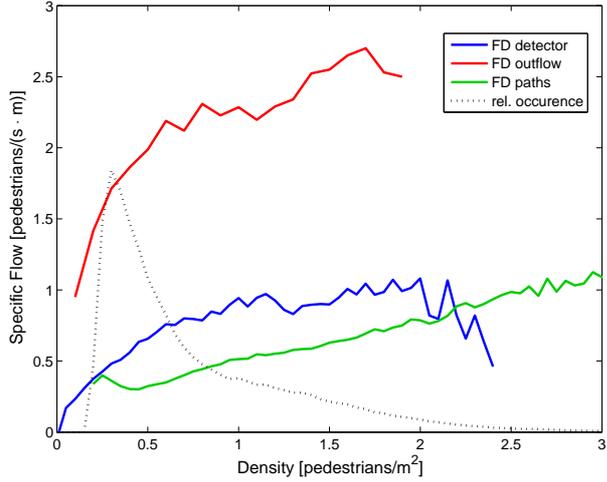}
\caption{Comparison of FD $J(\rho)$ evaluated from detector data (blue), outflow data (red) and pedestrians' path data (green). Dotted line illustrates the frequency of density's occurrence.}
\label{fig:srov FD} 
\end{figure}

The pedestrian behaviour in front of the exit have been analyzed by FD $J^{s}(\rho)$ (in Figure \ref{fig:srov FD}). The outflow, detector and pedestrian data have been compared.

The outflow data are measured just at the door, on contrary, the detector is placed in front of the door. The width of crossection, which is crossed by pedestrians, is 2 m. Due to the conservation law, the absolute flow must be conserved at all crossections which are used by all pedestrians. Therefore the specific flow at the door is much higher than the specific flow in the detector.

However, the trends of area and outflow data in FD are similar, this curves differs only in scaling. The flow linearly increases until the density reaches 0.3 (resp. 0.4) ped/m$^{2}$. This linear part characterizes the free flow state. Then, the flow continues to increase slowly and at the density 0.8 (resp. 0.9) ped/m$^{2}$, the flow is stabilized, the cluster is formed. The maximal flow occurs at the density 1.7 ped/m$^{2}$, where the second increase comes. At the densities larger than 2 ped/m$^{2}$, fast decrease occurs.

Using hydrodynamic relation (1), the FD $J(\rho)$ can be derived also from the pedestrians data. Surprisingly the trend is quite different. The pedestrians' velocity decreases rapidly when the density reaches 0.3 ped/m$^{2}$ and then it fluctuates around constant value (see Fig. \ref{fig:3DFD}). Thus the flow falls first and then increases linearly.

To conclude, the 3D FD generated by pedestrians' data clearly describes the phase of the system or the state of a pedestrian. But to provide the $J(\rho)$ FD, area based methods produce more relevant information.

All observation of the phase transition illustrates, that pedestrians change their velocity in advance, the slowdown process starts around 3 m in front of the obstacle.

From Table \ref{tab:prub_exp} it is clear that the assumption that phase transition is determined by the inflow parameter is correct. The cluster formation in front of the bottleneck is very sensitive to this parameter. The transition has been observed while the inflow is between  1.4 and 1.7 ped/s respectively between 1.7 and 1.9 ped/s depending on the size of the room.\\[1cm]

{\bf Acknowledgement:} This work was supported by the grant SGS12/197/OHK4/3T/14 and the research program MSM 6840770039.

%
%

\begin{thebibliography}{99.}%
%
%
%
%
%


\bibitem{Boltes2011} Boltes M., Zhang J., Seyfried A., Steffen B. T-junction: Experiments, trajectory collection, and analysis. ICCV Proceedings, 158--165 (2011)

\bibitem{Bakstein2006} Bakstein H., Havlena M. at all: Omnidirectional Sensors and Their Calibration for the Dirac Project, Research Reports on CMP 10, Czech Technical University in Prague (2006)

\bibitem{Ezaki2012} Ezaki T., Yanagisawa D.: Metastability in Pedestrian Evacuation. LNCS \textbf{7495}, 776--784 (2012)

\bibitem{Ezaki2012a} T. Ezaki at all, Simulation of Space Acquision Process of Pedestrians Using Proxemic Floor Field Model, Physica A \textbf{391}, 291--299 (2012)

\bibitem{Federici2012} Federici M. L. at all: Data Collection for Modeling and Simulation: Case Study at the University of Milan-Bicocca, LNCS \textbf{7495}, 699--708 (2012)

\bibitem{Gonzales2001} Gonzales R. C., Woods R. E.: Digital Image Processing, Prentice Hall (2001)

\bibitem{Georgoudas2011} Georgoudas I. G., Sirakoulis G. Ch. and  Andreadis I. Th.: An Anticipative Crowd Management System Preventing Clogging in Exits During Pedestrian Evacuation Processes, IEEE System Journal 2010 \textbf{5}, 129--141 (2011)

\bibitem{Helbing2000} Helbing D., Farkas I., Vicsek T.: Simulating Dynamical Features of Escape Panic, Nature \textbf{407}, 487--490 (2000)

\bibitem{Hrabak2012} Hrabak P., Bukacek M., Krbalek M.: Cellular Model of Room Evacuation Based on Occupancy and Movement Prediction, LNCS \textbf{7495}, 709--718 (2012)

\bibitem{Jelic2012} Jelic A., Appert-Rolland C. at all: Properties of Pedestrians Walking in Line -- Fundamental Diagrams, Phys. Rev. E \textbf{85/3}, 057302 (2012)

\bibitem{Plaue2011} Plaue M., Chen M. at all: Trajectory Extraction and Density Analysis of Intersecting Pedestrian Flows from Video Recordings. LNCS \textbf{6952}, 285--296 (2011)

\bibitem{Seyfried2010} Seyfried A., Boltes M. at all: Enhanced Empirical Data for the Fundamental Diagram and the Flow Through Bottlenecks, Pedestrian and Evacuation Dynamics 2008, 145-156 (2010)

\bibitem{Seyfried2010a} Seyfried A., Portz A., Schadschneider A.: Phase Coexistence in Congested States of Pedestrian Dynamics, LNCS \textbf{6350}, 496–505 (2010)

\bibitem{Schadschneider2010} Schadschneider A., Chowdhury D., Nishinari K.: Stochastic Transport in Complex Systems, Elsevier (2010)

\bibitem{Schadschneider2009} Schadschneider A., Seyfried A.: Empirical Results for Pedestrian Dynamics and their Implication for Cellular Automata Models, in Pedestrian Behavior: Models, Data Collection and Applications, Emerald Group Publishing (2009)

\bibitem{Steffen2010} Steffen B., Seyfried A.: Methods for Measuring Pedestrian Density, Flow, Speed and Direction with Minimal Scatter, Physica A \textbf{389(9)}, 1902--1910 (2010)

\bibitem{Was2011} Was J., Mysliwiec W., Lubas R.: Towards Realistic Modeling of Crowd Compressibility, Pedestrian and Evacuation Dynamics 2010, 527--534 (2011)

\bibitem{Was2010} Was J.: Experiments on Evacuation Dynamics for Different Classes of Situations, Pedestrian and Evacuation Dynamics 2008, 225--232 (2010)

\bibitem{Zhang2010} Zhang J., Klingsch W., Seyfried A.: High Precision Analysis of Unidirectional Pedestrian Flow within the Hermes Project. In: The Fifth Performance-based Fire Protection and Fire Protection Engineering Seminars (2010)

\end{thebibliography}
%

\end{document}